\documentclass[sigconf, screen]{acmart}

\usepackage{capt-of}
\usepackage{balance,bm}
\usepackage{color, colortbl, xcolor}

\usepackage [english]{babel}
\usepackage{url}

\usepackage{graphicx}  
\usepackage{textcomp, soul}
\usepackage{wrapfig}
\usepackage{subcaption}
\usepackage{color,soul}
\usepackage[ruled]{algorithm2e}
\usepackage[T1]{fontenc}
\usepackage[utf8]{inputenc}
\usepackage{microtype}
\usepackage{siunitx}
\usepackage{multirow}
\usepackage{enumitem}

\mathchardef\mhyphen="2D 

\usepackage [autostyle, english = american]{csquotes}
\definecolor{linkColor}{RGB}{6,125,233}
\definecolor{green}{rgb}{0.0, 0.65, 0.31}
\definecolor{bleudefrance}{rgb}{0.19, 0.55, 0.91}
\definecolor{ceruleanblue}{rgb}{0.16, 0.32, 0.75}
\definecolor{grey}{HTML}{969696}
\definecolor{violet}{HTML}{8856a7}
\definecolor{dgrey}{HTML}{01665e}
\definecolor{lgrey}{HTML}{5ab4ac}
\definecolor{dgreen}{HTML}{005a32}
\definecolor{purple}{HTML}{ae017e}

\definecolor{editCol}{HTML}{0000FF}
\definecolor{editCol2}{rgb}{0.1, 0.1, 0.8}

\definecolor{maskCol}{HTML}{c51b7d}
\definecolor{lrColor}{HTML}{8856a7}
\definecolor{trColor}{HTML}{d01c8b}
\definecolor{ctColor}{HTML}{4dac26}
\definecolor{brickred}{HTML}{f03b20}
\definecolor{improveCol}{HTML}{253494}
\definecolor{worsenCol}{HTML}{d7191c}
\definecolor{neutralCol}{HTML}{dd1c77}
\definecolor{neutralGreen}{HTML}{31a354}
\definecolor{bleudefrance}{rgb}{0.19, 0.55, 0.91}  
\definecolor{mediumblue}{rgb}{0.0, 0.0, 0.8}
\definecolor{lgrey}{HTML}{5ab4ac}
\definecolor{dgreen}{HTML}{005a32}
\definecolor{purple}{HTML}{ae017e}
\definecolor{lblue}{HTML}{decbe4}
\definecolor{deepgrey}{HTML}{525252}
\definecolor{dslate}{HTML}{2F4F4F}
\definecolor{dolive}{HTML}{556B2F}
\definecolor{teal}{HTML}{388E8E}
\definecolor{NewBlue}{HTML}{1879ba}
\definecolor{DarkBlue}{HTML}{00008B}
\definecolor{darkgolden}{HTML}{CD950C}
\definecolor{maskCol}{HTML}{c51b7d}
\definecolor{lrColor}{HTML}{8856a7}
\definecolor{trColor}{HTML}{d01c8b}
\definecolor{ctColor}{HTML}{4dac26}
\definecolor{brickred}{HTML}{f03b20}
\definecolor{improveCol}{HTML}{253494}
\definecolor{worsenCol}{HTML}{d7191c}
\definecolor{lgreen}{HTML}{e0f3db}
\definecolor{dpink}{HTML}{CD1076}
\definecolor{pink}{HTML}{FED2D2}
\definecolor{soothinggreen}{HTML}{4dac26}
\definecolor{darkred}{HTML}{8B0000}

\definecolor{mscolor}{HTML}{01665e}
\definecolor{nmscolor}{HTML}{bf812d}
\definecolor{lgreen}{HTML}{ccece6}

\usepackage{mathtools}

\usepackage{xcolor}
\usepackage{arydshln}
\makeatletter
\def\adl@drawiv#1#2#3{%
        \hskip.5\tabcolsep
        \xleaders#3{#2.5\@tempdimb #1{1}#2.5\@tempdimb}%
                #2\z@ plus1fil minus1fil\relax
        \hskip.5\tabcolsep}
\newcommand{\cdashlinelr}[1]{%
  \noalign{\vskip\aboverulesep
           \global\let\@dashdrawstore\adl@draw
           \global\let\adl@draw\adl@drawiv}
  \cdashline{#1}
  \noalign{\global\let\adl@draw\@dashdrawstore
           \vskip\belowrulesep}}
\makeatother

\newcommand*{\textlabel}[2]{%
  \edef\@currentlabel{#1}
  \phantomsection
  #1\label{#2}
}

\definecolor{policycolor}{HTML}{ffffb3}
\definecolor{purposecolor}{HTML}{e78ac3}
\definecolor{technologycolor}{HTML}{66c2a5}

\newcommand{\hlPur}[1]{\colorbox{purposecolor!20}{#1}}
\newcommand{\hlPol}[1]{\colorbox{policycolor!40}{#1}}

\newcommand{\cellPol}[1]{\cellcolor{policycolor!40}{#1}}
\newcommand{\cellPur}[1]{\cellcolor{purposecolor!20}{#1}}

\colorlet{tableheadcolor}{gray!25} 
\colorlet{tablerowcolor}{gray!15} 
\colorlet{tablerowcolor2}{gray!12} 
\colorlet{tablerowcolor3}{gray!25} 

\newcommand{\rowcollight}{\rowcolor{tablerowcolor2}} %

\renewcommand{\textrightarrow}{$\rightarrow$}

\renewcommand{\textrightarrow}{$\rightarrow$}

\colorlet{tableheadcolor}{gray!25} 
\colorlet{tablerowcolor}{gray!5} 

\definecolor{neutralCol}{HTML}{dd1c77}
\definecolor{neutralGreen}{HTML}{31a354}
\definecolor{NewBlue}{HTML}{1879ba}
\definecolor{bleudefrance}{rgb}{0.19, 0.55, 0.91}  
\definecolor{AfTrColor}{HTML}{0868ac}  
\definecolor{BfTrColor}{HTML}{a8ddb5}  

\definecolor{AfCtColor}{HTML}{b10026}  
\definecolor{BfCtColor}{HTML}{fd8d3c}

\graphicspath{ {figures/} }

 \definecolor{TrmtColor}{RGB}{47, 79, 79}
\definecolor{CtrlColor}{RGB}{162, 205, 90}
\definecolor{edit}{rgb}{0.0, 0.0, 0.0}

\newif{\ifhidecomments}
 \hidecommentsfalse 
\ifhidecomments
    \newcommand{\shreya}[1]{}
    \newcommand{\anna}[1]{}
    \newcommand{\koustuv}[1]{}
    \newcommand{\alexandra}[1]{}
    \newcommand{\mary}[1]{}
    \newcommand{\jina}[1]{}
\else
    \newcommand{\shreya}[1]{\textbf{\sffamily{\textcolor{DarkBlue}{[#1 -- Shreya]}}}}
    \newcommand{\anna}[1]{\textbf{\sffamily{\textcolor{violet}{[#1 -- Anna]}}}}
    \newcommand{\koustuv}[1]{\textbf{\sffamily{\textcolor{dpink}{[#1 -- Koustuv]}}}}
    \newcommand{\alexandra}[1]{\textbf{\sffamily{\textcolor{darkgolden}{[#1 -- Alexandra]}}}}
    \newcommand{\mary}[1]{\textbf{\sffamily{\textcolor{teal}{[#1 -- Mary]}}}}
    \newcommand{\jina}[1]{\textbf{\sffamily{\textcolor{dgreen}{[#1 -- Jina]}}}}
\fi

\newcommand{\para}[1]{\smallskip\noindent\textbf{#1}~}
\newcommand{\itpara}[1]{\smallskip\noindent\textit{#1}~}
\usepackage{acmart-taps}

\AtBeginDocument{%
  \providecommand\BibTeX{{%
    \normalfont B\kern-0.5em{\scshape i\kern-0.25em b}\kern-0.8em\TeX}}}

\setcopyright{none}

\acmConference[FAccT '23]{2023 ACM Conference on Fairness, Accountability, and Transparency}{June 12--15, 2023}{Chicago, IL, USA}
\acmBooktitle{2023 ACM Conference on Fairness, Accountability, and
Transparency (FAccT '23), June 12--15, 2023, Chicago, IL, USA}
\acmDOI{10.1145/3593013.3594023}
\acmISBN{979-8-4007-0192-4/23/06}

\begin{document}

\title[Can Workers Meaningfully Consent to Workplace Wellbeing Technologies?]{Can Workers Meaningfully Consent\\ to Workplace Wellbeing Technologies?}

\author{Shreya Chowdhary}
  \orcid{0009-0004-1799-7836}
\affiliation{%
  \institution{University of Michigan}
  \city{Ann Arbor}
  \state{MI}
  \country{USA}
}
\email{schowdha@umich.edu}

\author{Anna Kawakami}
\orcid{0009-0000-6937-6370}
\affiliation{%
  \institution{Carnegie Mellon University}
  \city{Pittsburgh}
  \state{PA}
  \country{USA}
}
\email{akawakam@andrew.cmu.edu}

\author{Mary L. Gray}
\orcid{0000-0001-9972-6829}
\affiliation{%
  \institution{Microsoft Research}
  \city{Cambridge}
  \state{MA}
  \country{USA}
}
\email{mlg@microsoft.com}

\author{Jina Suh}
\orcid{0000-0002-7646-5563}
\affiliation{%
  \institution{Microsoft Research}
  \city{Redmond}
  \state{WA}
  \country{USA}
}
\email{jinsuh@microsoft.com}

\author{Alexandra Olteanu}
\orcid{0000-0001-5710-4511}
\affiliation{%
  \institution{Microsoft Research}
  \city{Montreal}
  \state{QC}
  \country{Canada}
}
\email{alexandra.olteanu@microsoft.com}

\author{Koustuv Saha}
\orcid{0000-0002-8872-2934}
\affiliation{%
  \institution{University of Illinois at Urbana-Champaign}
  \city{Urbana}
  \state{IL}
  \country{USA}
}
\email{koustuv.saha@gmail.com}

\renewcommand{\shortauthors}{Chowdhary et al.}


\begin{abstract}

Sensing technologies deployed in the workplace can unobtrusively collect detailed data about individual activities and group interactions that are otherwise difficult to capture. A hopeful application of these technologies is that they can help businesses and workers optimize productivity and wellbeing. However, given the inherent and structural power dynamics in the workplace, the prevalent approach of accepting tacit compliance to monitor work activities rather than seeking workers' meaningful consent raises privacy and ethical concerns. This paper unpacks challenges workers face when consenting to workplace wellbeing technologies. Using a hypothetical case to prompt reflection among six multi-stakeholder focus groups involving 15 participants, we explored participants' expectations and capacity to consent to these technologies. We sketched possible interventions that could better support meaningful consent to workplace wellbeing technologies, by drawing on critical computing and feminist scholarship---which reframes consent from a purely individual choice to a structural condition experienced at the individual level that needs to be freely given, reversible, informed, enthusiastic, and specific (FRIES). 
The focus groups revealed how workers are vulnerable to ``meaningless'' consent---as they may be subject to power dynamics that minimize their ability to withhold consent and may thus experience an erosion of autonomy in their workplace, also undermining the value of data gathered in the name of ``wellbeing.''
To meaningfully consent, participants wanted changes to how the technology works and is being used, as well as to the policies and practices surrounding the technology. Our mapping of what prevents workers from meaningfully consenting to workplace wellbeing technologies (challenges) and what they require to do so (interventions) illustrates how the lack of meaningful consent is a structural problem requiring socio-technical solutions.

\end{abstract}


%
%



\maketitle



\section{Introduction~\label{Introduction}}
The evolving nature of the workplace, especially since the pandemic, has seen not only shifts towards remote and hybrid work~\cite{mark2022introduction,yang2022effects} but also efforts to integrate worker productivity and wellbeing support technologies into the workplace~\cite{dasswain2020social,litchfield2021workplace}.
These technologies often rely on passive sensing to enable unobtrusive collection of large-scale behavioral data in real-time and longitudinally, which can be used to infer complex insights about individuals such as mood and cognition~\cite{larradet2020toward,mark2016email,murali2021affectivespotlight,zakaria2019stressmon}, productivity and job performance~\cite{kaur2020optimizing,mirjafari2019differentiating}, or organizational fit~\cite{dasswain2019multisensor,DasSwain2019FitRoutine}. 
However, these technologies raise ethical and privacy concerns~\cite{corvite2022data,kaur2022didn,park2021human,abebe2020roles,holten2021can}. These are situated within a broader historical context of ``Taylorism'' and technology-enabled workplace surveillance~\cite{o2017taylorism}. While workplace surveillance was previously limited by time and resource costs, data-driven technologies have reduced such costs, expanding the possibilities of workplace surveillance and making it virtually limitless~\cite{ajunwa2017limitless,zickuhr2021workplace}. Even when the stated intention is to support workers, these technologies can still make workers feel controlled~\cite{mantello2021bosses, chung2017finding}. 

A key concern is workers' ability to consent to these technologies.~\citeauthor{willborn2005consenting} described consent as ``a crucial component of privacy that empowers individuals and affirms human dignity''~\cite{willborn2005consenting}. However, the combination of structural power dynamics within the workplace and the inherent intrusiveness of passive sensing technologies raise questions about whether workers can truly, meaningfully consent to these technologies, which is often assumed or compromised~\cite{moore2000employee,willborn2005consenting,palm2009securing}. That is, \textit{when an employer ``asks for'' a worker's consent, are they fully empowered to consent?} 
The current standard practice is informed consent, which involves informing users of the terms of consent and providing them with a binary yes/no choice to consent before their first use of the technology. Privacy research has however demonstrated that such consent practices may not meaningfully model consent
\cite{barocas2014big,solove2012introduction}, and are ill-suited for sensing technologies designed to unobtrusively collect large quantities of individual and contextual data~\cite{barocas2014big,luger2013consent}.  

Consequently, workplace wellbeing sensing technologies necessitate an approach to consent that can accommodate both the structural factors of the workplace and the complexities of the technology. Critical privacy research proposes de-prioritizing efforts to reform the procedural elements of consent and focusing on providing the legal, technical, and organizational conditions for consent~\cite{solove2012introduction, barocas2014big, nissenbaum2009privacy}. Critical computing and feminist scholarship reframed consent as a sociotechnical condition experienced at an individual level, as opposed to an individual choice~\cite{cohen2019turning}, and shifted the definition of consent from ``no means no'' to ``yes means yes''---a model that understands consent as contingent on one's relative agency and power~\cite{friedman2019yes,curtis2017affirmative}. \textit{A well-grounded means to define meaningful consent is when consent is freely-given, reversible, informed, enthusiastic, and specific, known as FRIES~\cite{parenthood2020sexual}.}
Guided by recent HCI research theorizing how to apply feminist consent models to technology~\cite{lee2017consentfultech, im2021yes, strengers2021embodied, varon2021artificial}, this paper conducts an empirical study of what it means for workers to provide meaningful consent to workplace wellbeing technologies.
In particular, we ask:
\begin{enumerate}
    \item[\textbf{RQ1:}] What are the challenges to meaningfully consent to wellbeing sensing technologies in the workplace?
    \item[\textbf{RQ2:}] What sociotechnical interventions can ameliorate these challenges and support meaningful consent? 
\end{enumerate}

We conducted six focus groups with 15 participants with diverse perspectives on consent and privacy, the workplace, and sensing technology. We centered the focus group discussions around FRIES~\cite{parenthood2020sexual} and sketches of sociotechnical interventions derived from prior work that might better support consent. We thematically analyzed participants' perspectives on how the workplace, technology, or both can produce meaningless consent and synthesized four layers of challenges: inherent power differential in the workplace; consequences of this power differential; inherent risks with the technology; and technological barriers. Our participants also suggested
interventions to change the technology's aims and affordances, as well as the policies and practices surrounding the technology. Drawing on these findings, we discuss implications for context-specific affordances necessary for meaningful consent to workplace wellbeing sensing technologies, the necessity (and difficulty) to trust an employer and how to foster this trust, and whether meaningful consent to these technologies is actually possible.

\para{Ethics \& Positionality.} 
Given our study's socio-organizational context, we explicitly assured the participants during the study that their participation was voluntary, all questions were optional, their responses would be anonymous, and they may ask us to remove any recorded content from the study. As focus groups may make participants privy to others' personal life experiences, we asked them not to share others' information. To protect participants' privacy, we analyzed the data in a de-identified fashion and paraphrased their quotes here. Our team comprises researchers holding diverse gender, racial, and cultural backgrounds, including people of color and immigrants, and holds interdisciplinary research expertise in the areas of HCI, UbiComp, CSCW, and critical computing. This study was approved by the IRB at Microsoft Research.
\section{Related Work}\label{section:rw}

{\bf The Quantified Workplace, Wellbeing Sensing, \& Workplace Surveillance.} The term ``Quantified Workplace'' was coined to describe the adoption of passive sensing technologies to quantify organizational dynamics, and facilitate individual self-reflection and positive behavioral change~\cite{mashhadi2016let}. Scholars have experimented with multimodal passive sensing in the workplace through smartphone, wearable, Bluetooth, and wireless sensors, and social interactions to infer job performance~\cite{mirjafari2019differentiating}, mood and cognition~\cite{binmorshed2019mood,mark2014capturing,schaule2018employing,binmorshed2022advancing,mark2016email,robles2021jointly}, organizational fit~\cite{DasSwain2019FitRoutine,dasswain2019multisensor}, engagement~\cite{mitra2017spread,rachuri2011sociablesense,dasswain2023focused,saha2023focus,saha2021job}, productivity~\cite{epstein2016taking,cambo2017breaksense}, organizational role~\cite{saha2019libra,nepal2020detecting}, and so on. For instance,~\citeauthor{binmorshed2022advancing} built machine learning models on passively sensed behavioral data such as keyboards, webcams, calendars, and email activity to measure worker stress~\cite{binmorshed2022advancing}. This body of research leverages the ability of passive sensing to unobtrusively collect in-the-moment behavioral data both longitudinally and in real-time, which enables overcoming some limitations of traditional survey-based measurements of wellbeing, such as recall bias, subjectivity, and compliance~\cite{scollon2009experience,tourangeau2000psychology}. However, these technologies raise several privacy and ethical concerns~\cite{gorm2016sharing,adler2022burnout,roemmich2021data,dasswain2023algorithmic}. Prior work has noted that sensing-driven measurements of wellbeing constructs suffer limitations such as data and measurement biases, poor proxies, unintended consequences~\cite{jacobs2021measurement,olteanu2019social,passi2019problem} and mismatches in self-perceived versus passively sensed wellbeing~\cite{das2022semantic,roemmich2021data,kaur2022didn}. 

These technologies have also been situated within a long history of workplace surveillance~\cite{ajunwa2017limitless} as a contemporary iteration of Taylorism~\cite{o2017taylorism,mantello2021bosses,moore2016quantified}---quantifying workers to maximize their productivity~\cite{littler1978understanding}. \citeauthor{maltseva2020wearables} identified possible consequences of wearables for workplace wellbeing, including dehumanizing and objectifying workers by turning them into ``resources'' to be utilized and optimized~\cite{maltseva2020wearables}. Others show how these supposedly wellbeing-supporting technologies can cause marginalized people to bear more emotional labor~\cite{latorre2019definition, stark2020don, mantello2021bosses}. Recently,~\citeauthor{constantinides2022good} studied factors concerning employees using these technologies in the workplace~\cite{constantinides2022good}, and~\citeauthor{kawakami2023wellbeing} noted the multi-layered nature of their harmful and beneficial impacts in terms of organizational, interpersonal, and individual-level impacts~\cite{kawakami2023wellbeing}.

The use of workplace technologies is situated amid complex power imbalance and workplace dynamics. 
Therefore, when employers ask workers to ``choose'' to consent to technologies that collect sensitive data and can compromise their privacy, \textit{consent may not be a choice}. Our work studies the challenges in the workplace that lead to meaningless consent and possible interventions to support meaningful consent to workplace wellbeing technologies. 

\para{Privacy, Consent, \& Sensing Tech.} The standard approach to privacy is privacy self-management~\cite{solove2012introduction}, or policies to help individuals manage their data. This includes notice and consent, or informed consent, which is grounded in the idea that if people are appropriately informed about the data collection and possible risks, then their consent is more legitimate as they understand and act out of free will. However, this model has been shown to produce meaningless consent as consent notices are often either too vague~\cite{solove2012introduction} or too complex~\cite{luger2013terms, bailey2021disclosures, custers2014privacy}, being over-prompted for consent can lead to ``consent fatigue''~\cite{schermer2014crisis}, and the ``choice'' is often meaningless~\cite{kovacs2020informed}.

The notice and consent model is ill-equipped to handle the privacy concerns raised by sensing technologies~\cite{barocas2014big,bird2016exploring}. Sensing technologies can unobtrusively collect user data; this inherent invisibility makes it difficult to appropriately notify and prompt users~\cite{luger2013consent}. Because these technologies rely on data to function, opting out while retaining technology access is virtually impossible~\cite{luger2014framework}. Data is often collected beyond an individual, making an individualized approach to consent ineffective~\cite{barocas2014big}. \citeauthor{luger2014framework} criticized standard consenting practices for sensing technologies for prioritizing ``security consent'' as opposed to ``user agency'' and proposed a theoretical framework for informed consent~\cite{luger2014framework}. Others have proposed technical re-designs to better support informed consent to sensing technologies through visualizations of information trajectories and system consequences~\cite{moran2014toolkit}, semi-autonomous consenting agents~\cite{gomer2014sac,jones2018ai}, learning privacy preferences over time~\cite{das2018personalized}, AI-powered chatbot~\cite{xiao2023inform}, and multi-layered privacy notices to actively engage users in the consent process~\cite{jones2018ai}.

\para{Consent as a Sociotechnical Condition.} To mitigate the limitations of informed consent, critical privacy scholars have proposed more systemic approaches that do not solely focus on procedural elements, like optimizing consent interfaces, but take a more infrastructural view of consent. They have conceptualized consent as a sociotechnical condition experienced at the individual level~\cite{cohen2019turning}, arguing for contextualizing consent within a bigger matrix of rights and obligations based on data sensitivity and the context~\cite{nissenbaum2009privacy, barocas2014big}.

Feminist scholarship, in particular, approaches consent with a focus on power and context~\cite{fahs2016sex,humphreys2007sexual}. Feminist and sex-positive scholars pioneered a movement to redefine consent from ``no means no'' to ``yes means yes'', or affirmative consent, recognizing that the ``no means no'' model ignores structural factors like gender norms making it difficult to say no~\cite{friedman2019yes,curtis2017affirmative}. Affirmative consent, in contrast, prioritizes individual agency over our bodies by honoring people's desires. Planned Parenthood breaks affirmative consent into five criteria, or \textbf{FRIES}~\cite{parenthood2020sexual}: 1) \textit{Freely given}: consent is a choice made without manipulation or influence; 2) \textit{Reversible}: consent can be revoked at any time; 3) \textit{Informed}: consent requires full comprehension of what one is consenting to; 4) \textit{Enthusiastic}: consent should be enthusiastically given; 5) \textit{Specific}: consent should not be taken as a monolith---consent to one thing does not mean consent to another.

Prior work has operationalized FRIES to design ``consentful technologies,'' where consent underlies the technology's design, development, data security, and user interaction~\cite{lee2017consentfultech}.
Recent HCI research has theorized affirmative consent as a way to preserve agency over ``data bodies''~\cite{im2021yes, strengers2021embodied, nguyen2020challenges,human2021human}, including on social platforms~\cite{im2021yes} and for technologies with embodied interactions~\cite{strengers2021embodied,nguyen2020challenges}. In addition, research has taken a feminist lens arguing for a more power-sensitive approach to consent to technology~\cite{pena2019consent,lee2017consentfultech}. \citeauthor{kovacs2020informed} reframed the consenting process by reconceptualizing the nature of data and centering bodies in the debates on data governance. 

However, consent to passive sensing tech largely remains a static process, while more meaningful consent to workplace wellbeing sensing technologies remains underexplored. Our work adopts FRIES as a starting point to ground a definition of meaningful consent that is sensitive to power asymmetries, like those that exist in the workplace, and inherently applicable to embodied interactions (as FRIES is constructed around sexual consent). Drawing inspiration from prior work on exploring FRIES for embodied interactions with technology~\cite{strengers2021embodied}, there is a precedent for adopting FRIES as a definition for meaningful consent for technologies that interact with our bodies, like wellbeing sensing technologies. By recasting consent as a sociotechnical condition, our work suggests interventions that recognize the power differential in the workplace and the technological complexities of passive sensing technologies.
\section{Study and Methods}

\subsection{Participants and Recruitment}

We aimed to recruit participants with varying perspectives about consent, the workplace, and passive sensing, including managers, legal professionals and policymakers, AI builders (developers, designers, researchers), and people with experience negotiating privacy in the context of a power asymmetry (i.e., union organizers, patient advocates). We recruited participants in August 2022 using a snowball and convenience sampling strategy by posting on organizational email lists and directly contacting people in different roles through emails and direct messaging. Interested respondents were provided with a consent form and responded to an in-take demographic survey (\autoref{table:participants}). 
We had 15 participants from 10 organizations of varying sizes, types (technology, consulting, research-based, non-profit), and four global locations (U.S., U.K., Canada, and India). Participants were compensated with \$50 USD Visa gift cards.

\aptLtoX[graphic=no,type=env]{\begin{table*}
    \caption{Overview of participants, with participant id (P\#), focus group (FG\#), country, demographics, organizational role, organization size, \& tenure in the organization. (* FG6 was technically a one-participant interview due to last-minute rescheduling.)}
    \begin{tabular}{lllllllll}
     \textbf{P\#} & \textbf{FG\#} & \textbf{Country} & \textbf{Gender} & \textbf{Race} & \textbf{Age} & \textbf{Role} & \textbf{Org.} & \textbf{Years}\\
     \toprule
    P1 & FG1 & US & Man & Asian & 25-34 & Researcher & - & 5-7 Ys.\\
    P2 & FG1 & U.S. & Woman & White & 25-34 & Program Manager/Legal & 10K+ & 1-3 Ys.\\
    P3 & FG1 & Canada & Man & Asian & 25-34 & Software Developer & 10K+ & 1-3 Ys.\\
    \hdashline
    P4 & FG2 & India & Woman & Asian & 25-34 & Policy Officer & 1 to 49 & 3-5 Ys.\\
    P5 & FG2 & Canada & - & - & 25-34 & AI Ethics Consultant & - & 1-3 Ys.\\
    \hdashline
    P6 & FG3 & U.S. & Woman & White & 35-44 & Content Designer & 50-1K & 3-5 Ys.\\
    P7 & FG3 & U.S. & Woman & Asian & 25-34 & Product Designer & - & 1-3 Ys.\\
    P8 & FG3 & U.S. & Woman & Asian & 35-44 & Design Researcher & 10K+ & <1 Yr.\\
    \hdashline
    P9 & FG4 & Canada & Man & White & 25-34 & PM/Editor/Researcher & 10K+ & 3-5 Ys.\\
    P10 & FG4 & Canada & Man & White & 35-44 & Researcher & 10K+ & 3-5 Ys.\\
    P11 & FG4 & U.S. & Woman & White & 25-34 & Researcher & - & <1 Yr.\\
    \hdashline
    P12 & FG5 & U.S. & Man & White & 25-34 & Researcher & 10K+ & 5-7 Ys.\\
    P13 & FG5 & U.S. & Woman & White & 25-34 & Software Engineer & 1K-5K & 5-7 Ys.\\
    P14 & FG5 & U.S. & Woman & White & 55-64 & Professor & 50-1K & 5-7 Ys.\\
    \hdashline
    P15 & FG6* & U.K. & Woman & White & 25-34 & Designer & 10K+ & 1-3 Ys.\\
    \\
    \bottomrule
    \end{tabular}\label{table:participants}
\end{table*}}{\begin{table*}[t]
\sffamily
\centering
\scriptsize
    \setlength{\tabcolsep}{1pt}
    \renewcommand{\arraystretch}{0.65}
  \centering
    \caption{Overview of participants, with participant id (P\#), focus group (FG\#), country, demographics, organizational role, organization size, \& tenure in the organization. (* FG6 was technically a one-participant interview due to last-minute rescheduling.)}
    \vspace{-6pt}
    \begin{minipage}[t]{\columnwidth}
    \resizebox{\columnwidth}{!}{
    \begin{tabular}{lllllllll}
     \textbf{P\#} & \textbf{FG\#} & \textbf{Country} & \textbf{Gender} & \textbf{Race} & \textbf{Age} & \textbf{Role} & \textbf{Org.} & \textbf{Years}\\
     \toprule
    P1 & FG1 & US & Man & Asian & 25-34 & Researcher & - & 5-7 Ys.\\
    P2 & FG1 & U.S. & Woman & White & 25-34 & Program Manager/Legal & 10K+ & 1-3 Ys.\\
    P3 & FG1 & Canada & Man & Asian & 25-34 & Software Developer & 10K+ & 1-3 Ys.\\
    \hdashline
    P4 & FG2 & India & Woman & Asian & 25-34 & Policy Officer & 1 to 49 & 3-5 Ys.\\
    P5 & FG2 & Canada & - & - & 25-34 & AI Ethics Consultant & - & 1-3 Ys.\\
    \hdashline
    P6 & FG3 & U.S. & Woman & White & 35-44 & Content Designer & 50-1K & 3-5 Ys.\\
    P7 & FG3 & U.S. & Woman & Asian & 25-34 & Product Designer & - & 1-3 Ys.\\
    P8 & FG3 & U.S. & Woman & Asian & 35-44 & Design Researcher & 10K+ & <1 Yr.\\
    \bottomrule
    \end{tabular}}
    \end{minipage}\hfill
        \begin{minipage}[t]{\columnwidth}
    \resizebox{\columnwidth}{!}{\begin{tabular}{lllllllll}
     \textbf{P\#} & \textbf{FG\#} & \textbf{Country} & \textbf{Gender} & \textbf{Race} & \textbf{Age} & \textbf{Role} & \textbf{Org.} & \textbf{Years}\\
     \toprule
    P9 & FG4 & Canada & Man & White & 25-34 & PM/Editor/Researcher & 10K+ & 3-5 Ys.\\
    P10 & FG4 & Canada & Man & White & 35-44 & Researcher & 10K+ & 3-5 Ys.\\
    P11 & FG4 & U.S. & Woman & White & 25-34 & Researcher & - & <1 Yr.\\
    \hdashline
    P12 & FG5 & U.S. & Man & White & 25-34 & Researcher & 10K+ & 5-7 Ys.\\
    P13 & FG5 & U.S. & Woman & White & 25-34 & Software Engineer & 1K-5K & 5-7 Ys.\\
    P14 & FG5 & U.S. & Woman & White & 55-64 & Professor & 50-1K & 5-7 Ys.\\
    \hdashline
    P15 & FG6* & U.K. & Woman & White & 25-34 & Designer & 10K+ & 1-3 Ys.\\
    \\
    \bottomrule
    \end{tabular}}
    \end{minipage}
\label{table:participants}
\vspace{-4pt}
\end{table*}}

\subsection{Study Design}

We conducted several focus groups (FG) with participants balanced across demographics, organizations, and roles to convene a diversity of multi-stakeholder conversations. The FGs were designed to be small and structured to allow different stakeholders to complement and supplement each others' perspectives on consent in the workplace. Our rationale for conducting FGs rests on the notion that FGs would enable participants to build off of each others' expertise and experience. We hypothesized that based on their roles, participants may notice different challenges and possible refinements to interventions, leading to more robust recommendations and a multi-layered, socio-technical approach to consent.

We conducted six $\sim$90-minute FGs via Microsoft Teams that we recorded with participants' consent. Each FG included three parts: 1) an information-gathering phase, 2) an overview of a hypothetical workplace wellbeing sensing technology, and 3) an intervention-refinement phase. We describe these below.

\para{Information Gathering Phase.}
The information-gathering phase elicited participants' perspectives on consent. After participants shared their general perspectives on consent (interpersonal, to technology, and/or in the workplace), we grounded the discussions on the FRIES consent model---under which meaningful consent needs to be \textbf{f}reely given, \textbf{r}eversible, \textbf{i}nformed, \textbf{e}nthusiastic, and \textbf{s}pecific~\cite{parenthood2020sexual}.
This provided a common definition for our participants to draw on while reflecting on past experiences with consenting to workplace technologies and a hypothetical worker wellbeing sensing technology we presented them with.

\para{Hypothetical Workplace Wellbeing Technology: \textit{Amellio}.}
After providing a brief overview of passive sensing technologies and how they function, we asked participants to imagine that their workplace, to support worker wellbeing, implemented Amellio---\textit{an app that passively collects your data using your work device and a wearable sensor, and provides insights to you about your wellbeing. You are not required to use Amellio, though managers encourage using it}. We then showed participants a sketch of Amellio's consent prompt (Appendix \autoref{fig:amellio_terms}). We conceptualized Amellio and designed this sketch by drawing on existing wellbeing sensing, as well as consent forms used by studies of workplace wellbeing through passive sensing~\cite{mattingly2019tesserae,wang2014studentlife,binmorshed2022advancing}. We chose to keep the sketch vague to avoid being overly prescriptive of what Amellio does, hoping to encourage a variety of perspectives in the focus groups.

\para{Intervention Refinement.}
In the intervention refinement phase, we sought participant perspectives on possible socio-technical interventions to support meaningful consent.
We drew on the prior work~\cite{im2021yes,strengers2021embodied,ajunwa2017limitless,maltseva2020wearables,raji2022fallacy} to make low-fidelity sketches of interventions on different aspects of the system (overviewed below).
Using these sketches as a starting point for conversations, we elicited participants' feedback on which elements of FRIES were supported by these interventions, what they would change, what issues they foresee with the implementation of these interventions, and what alternative or additional interventions they would suggest.  

    \itpara{-- Interface design change.} Inspired by~\cite{im2021yes, strengers2021embodied}, we presented participants with a sketch of a re-designed consent interface that included several improvements to better support consent (Appendix~\autoref{fig:intervention1}), including greater interactivity, improved visual design, more granular choices, more transparency about the system, and the ability to ask questions. 
    
    \itpara{-- Organizational practice.}
    Inspired by recommendations from \cite{maltseva2020wearables}, we presented participants with an outline for a boundary-setting conversation (Appendix~\autoref{fig:intervention2}) where they could speak with their manager and an Amellio developer to learn more about how it works and address concerns.
    
    \itpara{-- Policy and regulation.} Inspired by recommendations for regulatory stipulations, such as employee-health-specific HIPAA~\cite{ajunwa2017limitless}, we outlined a governmental policy (Appendix~\autoref{fig:intervention3}) that stipulates specific practices companies would need to comply with when handling employee health data.

\para{Data Analysis.}
After completing the focus groups,
we anonymized and qualitatively coded the interview transcripts. We began with line-by-line coding, followed by a thematic analysis to gradually coalesce the codes into relevant themes. Within each theme, we also categorized whether the code was more related to challenges or interventions. After iterative coding and categorizing the themes, we synthesized a set of both challenges and possible interventions.

\section{Results}
Our thematic analysis surfaced sociotechnical 
factors that lead to workers' meaningless consent to workplace wellbeing sensing technologies. In analyzing the data, we looked for participant perspectives on the \textbf{challenges} workers face (RQ1, \autoref{sec:challenges}), and the \textbf{interventions} to help mitigate these challenges (RQ2, \autoref{sec:interventions}).

\subsection{RQ1: Challenges to Meaningful Consent}\label{sec:challenges}

\begin{figure*}[t]
\centering
    \centering
    \includegraphics[width=2\columnwidth]{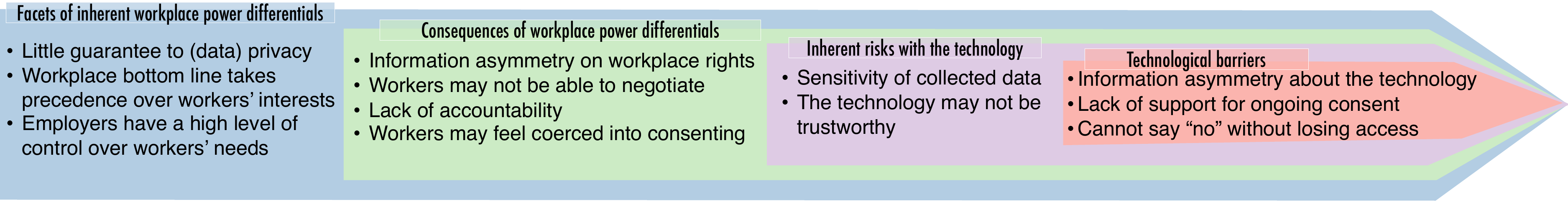}
    \vspace{-6pt}
    \caption{Themes related to challenges in consenting to workplace wellbeing sensing technologies.}
    \Description[figure]{A figure on the themes of layers of challenges to meaningful consent in the workplace, as surfaced in our focus groups. These layered themes in the increasing order of specificity to the technology are 1) facets of inherent workplace power differentials, 2) consequences of workplace power differentials, 3) inherent risks with the technology, and 4) technological barriers. Each of these layers include challenges that participants expressed that workers face when consenting to workplace wellbeing technologies.}   
    \label{fig:challenges}
    \vspace{-4pt}
\end{figure*}

We found four broad themes related to challenges: 1) facets of inherent power differential in the workplace, 2) consequences of the power differential, 3) inherent risks posed by the technology, and 4) technological barriers to supporting affirmative consent.~\autoref{fig:challenges} overviews these themes of challenges, which we discuss below. 

\subsubsection{Facets of inherent power differential in the workplace}
Participants described challenges to meaningful consent that were facets of an inherent power asymmetry in the workplace.

\para{Little guarantee to (data) privacy in the workplace.}
Participants noted workers are currently not guaranteed data privacy in the workplace. They pointed out employers are allowed to collect vast amounts of data on workers and are deemed owners of any data produced by employees, compromising consent and privacy. P14 expressed distrust about employers' ownership of workplace devices: 
``There's just an inherent distrust I have about the security of the data collected. The fact that it's collected at the workplace makes me think, who needs to know, and why? And if it is collected on devices installed by my employers, they are necessarily connected to the data storage. So these questions need to be addressed before someone can meaningfully consent to be part of this.''

Participants also thought the legal landscape that guarantees employers' ownership of work devices limits the effectiveness of strong data security practices, and the black box nature of these technologies could lead to surveillance workers may not be aware of; with P13 noting: ``I am concerned that I don't have a good read on what my company is doing otherwise to monitor my activity. I have no way of knowing if they have a keylogger on my machine, or if they have any sort of remote desktop watching stuff. Just because they have consent doesn't mean that it doesn't exist, and so, even if this is completely encrypted and secreted away and management has no insight into the data or anything after, they can still see if I click in [to Amellio], hypothetically. I cannot know that.''

\para{The workplace bottom line takes precedence over workers' interests.} 
Although Amellio is framed as supporting worker wellbeing, participants voiced their distrust e.g., ``If my manager is encouraging me to use this, my first question would be, why? What's the benefit for them and what's the visibility that they have over the system?''~[P15].
Participants with a labor rights background called attention to what they thought was employers' ulterior motive---reducing healthcare costs and boosting productivity to increase profit.
P12 described how employers often make decisions that ultimately benefit them, even when they claim to support workers: ``My employer once required me to install something that reminded me to look away from the screen and stop typing for a bit to avoid some eye disease or carpal tunnel syndrome. These protect the interests of the employer. I could never imagine a world where an employer says, `it looks like you're really stressed out, you should take advantage of the unlimited vacation;' there are certain things that are good for your health but against the employer's interests.''

\para{Employers have a high level of control over workers' needs.}
Participants noted how the high level control employers have over workers' needs makes consent to workplace technologies especially difficult.
P14 expressed concern about how workers are at risk because ``the power differential is related to your ability to make a living'' and expressed skepticism about whether this power differential can ultimately be mitigated at all. In addition, US workers rely on employers for health insurance, which makes sharing health data with employers especially risky. To navigate through this risk, P8 reframed consent as a relational process dependent on how trustworthy their employer was and how reliant they were on them, ``The chance that my employer will use [my data] to model eventually my [health] benefit, is a possibility I don't want to contribute to. That's where it is a relational thing. If I do trust my employer to do the ethical thing compared to some other company where I wouldn't feel that way, compared to working at a nonprofit where they don't actually give me much benefits anyways, so it's OK.''

\subsubsection{Consequences of the power differential}
Participants described other challenges to consent that could be characterized as consequences of the inherent power differential in the workplace.

\para{Information asymmetry about workplace rights and use of collected data.}
Participants remarked that workers may not know what their rights are in the workplace.
This is not only caused by the legal landscape that may not properly protect workers' rights, but also because employers may not transparently communicate those rights.
Drawing on their experience as a union organizer, P13 pointed out HIPAA, emphasizing that the lack of knowledge might prevent people from being able to make informed decisions: ``People are aware that HIPAA exists, that it protects them, but they don't understand anything about the context. They think it applies to their employer. Of course it doesn't.''

Participants thought that information asymmetry is also caused by the technology's black box nature.  They were concerned about the risks related to the data and the company's stewardship and use of the collected data. P5 asked ``[t]his data is going to fit which tasks? Who is responsible for that? What is the risk associated with this? What is the actual privacy action plan? Where can I find documentation to read more? How do you ensure the system is responsibly maintained over time?'' P5 also expressed a desire for greater transparency about how the technology is used.

\para{Workers may not be able to negotiate.}
Participants raised concerns about workers' inability to negotiate due to existing power dynamics and critiqued the idea of a boundary-setting conversation as the power difference between workers and managers makes true negotiation difficult. P11 noted the difficulty of being honest with managers: ``the people who are maybe uncomfortable to address these concerns to the manager or have no intention of actually using [Amellio], they would certainly not show up for such a conversation.'' For P13 the boundary-setting conversation with the manager is like ``outing yourself as someone with concerns.'' Participants also worried about penalties for voicing concerns.
P12 brought up the context of being neurodivergent: ``there are certain concerns I would love to tell my manager, but I don't really want [Amellio] to out me as autistic. I'm not gonna say that to my manager. If I have concerns with the system that is against the interest of the employer, I'm not going to say that to my manager.''

\para{Lack of accountability.}
Participants brought up accountability as a missing criterion for consent, arising from how employers own the technology and prioritize their interests. P13 expressed skepticism about the efficacy of the boundary-setting conversation because employers are not beholden to any legitimate responsibility to workers: ``I have no reason to believe that my manager has the answers to these, or is going to tell me the actual answer to this---he's not beholden to any sort of oath of truth here and can just say whatever wants.'' Participants also expressed apprehension about the potential misuse of their data and data-driven inferences. 
In fact, not only the original terms, but also the re-designed consent interfaces, could not help clarify P5's questions about whether the company owned the data and what their accountability processes looked like. Another element here is the lack of adequate legal means for accountability: ``The main way in which I've seen the company breaking the law is through unfair labor practices, a wide range of things where they're impinging on worker's right. This goes to the National Labor Relations Board, which is very slow and bureaucratic. There are very low stakes for the company and they can pay lawyers to beat down whoever else. [..] There would need of a rapid turn around and a very severe consequence''~[P13].

\para{Workers may feel coerced into consenting} or may not be given an actual \textit{choice} to consent given the employers' control over workers. Several participants described consent in the workplace as ``tacit'' or ``expected''. P2 attributed this to the
``overarching agreement between yourself and your employer'' which thus made consent less ``staggered or point-to-point''.
P1 similarly expressed that employment carries an expectation of agreeing and their consent could even be implicitly assumed. Others pointed out how this implicit expectation could manifest in different ways, even when there is increased specificity or space for negotiation. For example, 
after reviewing the re-designed consent interface, P14 noted valuable improvements but still felt ``there's an assumption from the beginning when I'm reading this, I'm going to say yes, so it already feels odd.'' Participants also felt that even though the boundary-setting conversation intended to create space for negotiation and reflection for the worker, it could feel coercive, as another effort to make the worker agree. 
P5, P8, P9, and P10 named the power imbalance between workers and managers as a source of potential coercion, expressing that workers may feel coerced into consenting out of a desire to not disobey or disappoint their managers (P10), or fear of damaging their relationship with the manager (P5). In contrast, other participants found it coercive only if workers were individually targeted but ``if this was just given out as a blanket thing to all employees, then I think that would be fine, compared to specifically targeting an individual''~[P3].
Additionally, participants described how workers may also feel coerced out of fear of e.g., losing their job: ``I think workplace have nuances like how much power the employee has. The Amazon warehouse workers probably wrote somewhere that they consent to wear that weird band. But the recourse is, well, then you won't work here''~[P12]. Another participant, P13 described experiences with technically being given a choice but they felt coerced to consent, because of penalties or that ``there might be a lot of consequences.''

\subsubsection{Inherent risks posed by the technology} Another set of challenges that emerged from our focus groups related to the inherent nature of passive sensing technologies.

\para{Passively collected data and data-driven inferences are sensitive and risky.}
Participants were concerned that the nature and amount of data collected are sensitive and could lead to negative consequences. P4 was concerned about how Amellio's use of a camera might also threaten the privacy of others in the workplace, while P2 noted that through passive sensing, previously ``non-sensitive'' data collected in aggregate could lead to sensitive inferences and require stricter regulation to protect privacy.
Other participants also noted that even the collection of seemingly non-sensitive and harmless data in large quantities could lead to other risks e.g., ``the metadata surrounding [Amellio], even if I am getting the feedback just for me, I don't have any faith about how the rest of it is being stored, how secured the system is, and I think it could lead to a penalty down the road, even if it is not the intent''~[P13].

\para{The technology itself may not be trustworthy (fair, unbiased, accurate \& beneficial).} Some participants,
especially AI builders, expressed concerns about the technology being fundamentally untrustworthy. P15 questioned the validity, ``I say this as someone who has some kind of understanding of AI. Maybe someone from a different background wouldn't have the same reaction. I simply think it is not possible to infer emotions from facial expressions. [..] there's a lot of gray areas with this type of technology and I don't really see the value of it either.'' Likewise, P5 questioned the construct validity and potential inaccuracies in Amellio's inferences: ``I want to know the assumptions and proxies used, because happiness, excitement, and distress are social constructs, and have many interpretations related to culture, for example. It's better to include some sort of clarification of how [Amellio] identifies them.''

Participants were concerned about potential biases and related consequences e.g., if the app mistakenly assesses them as always unhappy or unsatisfied (P11). P13 noted that being neurodivergent, they were concerned about the technology's reliability and benefits: ``as someone who's neurodivergent, things like this [technology] frequently doesn't work for me. [..] Is this likely gonna do more harm than good? My distractability or other things might be completely different from other people.'' Participants sought evidence about the technology's efficacy: ``I would like to see studies or some evidence backing up that tracking these kinds of data can actually improve employee health or wellbeing''~[P3].

\subsubsection{Technological challenges to supporting meaningful consent}\label{sec:technological_barriers}

These challenges relate to the design and mechanics of wellbeing sensing technologies, as well as to how people interact with them.

\para{Information asymmetry about the technology.}
Participants believe the lack of transparency about how the technology works and its limitations make it difficult to provide meaningful consent. After reviewing the re-designed consent interfaces, P14 critiqued the design for not providing adequate information: ``You're asking to check boxes to agree before the complete information is offered. We haven't described data analysis yet, so how can you consent? It's not true consent if we don't understand how the information could be used, and the risks also need to be described before they can truly consent.'' Similar concerns were raised about the boundary-setting conversations, with P8 emphasizing the lack of background knowledge as ``setting up a conversation between a developer assumes that you will be able to have a conversation,'' and P14 emphasizing that the ``uninformed don't even know what they don't know.''

\para{Technology may not support ongoing consent.}
Participants noted that sensing technologies do not typically support evolving or ongoing consent, i.e., consent should be an ongoing process~\cite{o2011ongoing}. They stressed how technologies often collect massive amounts of data in the background, but rarely support the ``reversible'' dimension of FRIES. P7 described how most services, after securing the first consent agreement, do not present an easy way to review, revise, or back out of the agreement: ``After I've given consent to something, it's always presented to you as the front door to enter [..] And then, after that point, it gets forgotten and there isn't a clear way to revise that consent or go back and be like, hey, I don't remember what the exact terms are.'' P2 was similarly concerned with sensing technologies being ill-equipped to handle ongoing consent over a long period of time as the technology itself evolves.

\para{Cannot say no without losing access to the technology.}
Participants felt unable to decline consent to only parts of the technology without losing all access. P6 pointed out how the initial terms of Amellio asked for consent as all-or-nothing and violated the ``specific'' dimension of FRIES: ``You are either agreeing to all of these data being collected or none of it, so it is definitely not specific consent.'' Multiple participants thought the re-designed consent interface better-supported specificity, especially the option to select which specific data they wanted to share. Those with AI building experience, however, were hesitant if this level of specificity could be technically supported. P10 contemplated the possibility of disentangling a specific data stream from the technology's accuracy, but found it somewhat impractical ``now that everything's basically a deep neural network where we don't understand the internals.'' Drawing on their expertise in developing passive sensing technologies, P1 also noted that decoupling the data sharing from the insights might not be possible as users ``can't be like, hey, I wanna see how my sleep looks, but I'm not enthusiastic about giving my data [to FitBit]--whose sleep data would you be looking at?''

Participants also thought this violates other FRIES dimensions, such as ``freely given'' and ``enthusiastic,'' as workers might feel they had no choice but to consent. Participants reflected on how the inextricability of ``good stuff'' and ``bad stuff'' creates a situation where the decision to consent becomes ultimately about trade-offs and not so much about enthusiasm. P6 noted how, under standard consent paradigms, they often find that enthusiasm was irrelevant and not supported at all: ``I really don't think [enthusiasm] is the norm for privacy consents [..] I know I don't really trust Facebook but I really value what I'm getting out of using Facebook. So it's like a trade-off.'' Similarly, P3 expressed, ``I feel sometimes you don't have to be enthusiastic. You could be more neutral about it, especially when wanting to use a certain service or product. Sometimes I just wanna use a product, and maybe I'm not enthusiastic about giving them my data and some personal information, but I do it as a trade-off for having that kind of access.'' 
\subsection{RQ2: Socio-technical Interventions to Support Meaningful Consent}\label{sec:interventions}
To organize intervention-related themes, we drew on the Consentful Tech Project~\cite{lee2017consentfultech} and mapped the interventions into two major themes: 1) \textit{technology's aims and affordances} and 2) \textit{policies and practices around the technology}.

\subsubsection{Technology's Aims and Affordances}
Our focus groups surfaced interventions that could help address concerns about \textit{why} and \textit{for whom} the technology was developed, i.e., its purpose and whose interests it prioritizes or serves. Many of these interventions recommend centering the workers' interests, as well as changing how the technology works, how it interacts with users, and what information about it is disclosed to workers.

\para{The technology should have a clear and sensible goal.} Participants interrogated the purpose of wellbeing sensing in the workplace, asking ``to what end'' the data was being collected, as they were suspicious of the employers' intentions and sought a clear purpose that actually benefits workers. These concerns can be mitigated by re-centering workers' interests as the driving motivator for developing such technologies. For P15 ``the most important thing would be introducing this to me as something to help me and help me grow in my job, rather than something helping managers manage their teams.'' However, others questioned the efficacy of simply stating the intended benefits. P10 related this insufficiency to workers' lack of trust in employers:
``[Communicating the benefits explicitly] could definitely drift into a creepier kind of coercive domain [..] even if they tell me there are benefits, do I believe them?''

\para{Provide evidence \& information about the technology's efficacy, limits, and risks.} Participants were skeptical about how well the technology worked, wanting evidence included in the consent agreement such as ``a study that looked through it, saying that for people using similar technology, e.g., depression rates went down or something, or certain behaviors improved''~[P3]. They also valued being able to speak to developers as a ``sign that they want to be transparent and want to gain your confidence''~[P3]. Participants further wished to know they can ``trust the software to do the right thing even without my oversight''~[P8] and sought an explicit ``ethics statement, describing the objectives of whoever’s collecting the data [and] their commitment to ethics''~[P10]. Participants sought more information on the limitations, tradeoffs, and impact assessment of the algorithm, and P1 drew analogies from nutrition labels in describing possible risks of the algorithm, ``like how you have a nutrition label and before purchasing, the person has the option to look at it.'' 

\para{The technology's benefits should be commensurate with the cost of data sharing.} P9 wanted to be treated ``like a shareholder [..] if I consent to give you my chat logs to improve the product, I want some share,''  adding they did not expect actual profits but something commensurate with the cost of sharing their data. They asserted that this could help re-balance whose interests are served by the system by making employers more responsible to the employees. Expanding on this idea, P10 described how being treated like a shareholder requires transparency, ongoing consent, and accountability for consequences. Participants want these systems to be responsible not only to workers but also to society. P11 noted that these technologies should only be used ``for making a positive difference in our society'' and that they would not be comfortable consenting if the data was used to harm others. P8 made a similar point that ``[i]s not just making sure that harm is not being caused, but that good intent is clearly stated, is being upheld.''

\para{The technology should facilitate functionality-based consent.} Participants were concerned about how the all-or-nothing consent model often makes them unable to refuse without losing complete access to the technology. Rather, mentioning the specific dimension of FRIES, they wanted access to only particular features of the technology that they were okay consenting with, either always or in specific scenarios. P1 drew on the example of applications like Signal~\cite{afzal2021encrypted}, which only requests camera access if the user wants to use the camera. 

\para{The technology should facilitate an active reflection and adjustment of consent preferences.} This will get closer to supporting ongoing consent and the reversible dimension of FRIES that a worker can revisit and adjust their consent preferences over time. P1 suggested preemptive mechanisms so that individuals know enough about their data at regular intervals, while P8 noted that ``the only time a service reaches out and asks me to change anything if there is a change in the agreement. They don't say [..] please read it again and refresh your memory. It would be good if it said, you've been using this service for five years; why don't you look at your privacy agreement?''

\para{Build interfaces that enable workers to better voice nuanced choices.} 
P13 observed ``[t]here's no clear way to say no. The assumption is that you'll just close the window, but I think that's not standard user behavior. They would rather click through next after unselecting or clicking no or something like that to be more explicit.'' P1 also challenged the common assumption that consent is a binary yes/no choice, and suggested a more granular scale, such as framing each dimension of FRIES as a question (e.g., how enthusiastic are you about this technology?).

\subsubsection{Policies and Practices Surrounding the Technology}
Our focus groups surfaced interventions that regulate the design and use of technology within the workplace context.

\para{Provide clarity about data governance practices.}
Participants sought greater clarity about data governance related to data retention and data minimization. P1 desired clarity over who owned the data and data-driven inferences and wanted to limit data collection to what is minimally needed, while P7 expected ``a lot more specific[s] about how long the data [is] stored. Is there a way to remove the data after a certain time; what happens after leaving the company?'' Similarly, P2 emphasized how the information on data handling is also important for accountability, ``[i]f an employer is using this technology, who’s responsible for the abuses and what if Amellio mishandles and abuses the data?''

\para{Be transparent about who has access to workers' data.}
Participants desired transparency and accountability from the employer, including explicit and ongoing communication about who can access their data as well as control over how and what data was shared. P8 sees such explicit communication as a sign of respect, noting how not having this information impacts their level of trust and comfort; while P9 suggested a ``weekly digest of all the information that is sent to others, how often the data is sent live, and if there are weekly trends'' to help them feel more comfortable.

\para{Show a strong record of trustworthy behavior.} Related to ``benevolent employer,'' participants suggested building a trustful relationship is essential for the employer or technology provider, to help foster trust and support meaningful consent. P11 noted that they would trust the technology more if they ``[knew] that this company has developed such technologies in the past and it has users' interests at heart,'' P8 mentioned employers demonstrating ``a high degree of ethical hygiene'' is a precondition for trust. Others, like P5 and P6, emphasized the relational nature of trust, dependent on an individual's relationship and experiences with the employer, while P15 noted that ``if I feel this company really cares for me, then I would be much more likely to get something like this.''

\para{Ensure oversight from a strong public interest organization.}
Participants thought an external body can enforce organizational commitments, ensure accountability, and foster trust, with P2 asking: ``can someone reasonably get their damages if their data was misused, or could a company reasonably be held accountable?'' Similarly, P10 noted ``[i]f it is just the company holding themselves accountable, I trust less than if there’s some external oversight that’s helping.'' P12 gave the example of the European Workers’ Council, which effectively creates barriers that prevent data abuse.

\para{Do not assume consent in the workplace.} 
To preserve workers’ autonomy, participants asked for practices that prevented coercion and preserve workers’ ability to say no. P9 wanted organizations to adopt an ``opt-in'' rather than ``opt-out'' policy. P4 and P8 underscored that preserving autonomy requires guarantees of no penalties or differential treatment for saying no. The decision to consent should also be personal. P12 stressed that others having knowledge of a worker’s consent decision could erode existing norms: ``I also really don’t think that it’s OK if your manager knows whether you signed up or no, because then that can change norms [and] can erode what is considered normal.''

\para{Prioritize the interests of workers above business.} Drawing on their experience, P14 discussed how a benevolent employer who centers workers' interests could mitigate the inherent power differential between workers and employers: ``I was the executive director of a nonprofit for families with children with brain tumors. I had volunteers [..] I can imagine a scenario where [I'd use the technology] because I would imagine myself to be a benevolent employer with their best interests at heart.'' P12 however critiqued this idea, suggesting that ``even if the employer is completely benevolent and wants to look after my health, there are ways that they can do that without centralizing power over the system [..] The employer can say, here's a \$500 benefit, go buy some health technology if you want, or sign up for a gym membership, we aren't gonna know what you do with it, but we care about you being healthy.'

\para{Create negotiation space centering workers' interests.} This would enable workers to negotiate more effectively while deterring organizations from prioritizing their bottom line. 
P2 appreciated the boundary-setting conversation as providing a negotiation space, as it is otherwise rare for users to interact with the technology developers and providers. In the re-designed consent interface, participants valued ``[t]he fact that I can ask questions if I am concerned[;] is a big plus that gives me more confidence''~[P11]. Participants also noted that a truly open space for negotiation is possible if the power differential is directly confronted and addressed: ``As long as [power differential] is addressed, understood, and agreed upon, you can move forward''~[P4].
To address the power differential, participants suggested group conversations between multiple workers and the manager and developer as a more comfortable and influential arrangement [P10], replacing the manager in the boundary-setting conversation with another non-supervising mentor [P1], and involving third parties who are invested in workers' interests only.

\para{Encourage deliberation in the consent process to preserve workers' autonomy.} P1 noted that ``[w]e typically think [of] consent procedures as very monolithic. It happens in one go, but maybe one way could be staggering it out, where you get some time to think over and come back and then make a decision.''
P7 similarly argued for more tailored approaches to the consent that meet each person's needs: ``As someone who is more introverted, I prefer reading in email or chat, and being able to sit with my thoughts, reflect, and come up with questions before engaging with someone as opposed to directly meeting them.'' P7 added that the space for deliberation may be further complemented by being able to ask additional context-related questions.

\para{Support recourse and enable workers to interrogate data practices.}
Our participants thought better recourse and interrogation opportunities would not only help realize the reversibility and specificity elements of FRIES, but also help enforce ethical commitments and hold employers accountable for any data abuse. For P8, ``being clear about ways to push back [and] understand when breaches will happen'' was essential to consent. P2 described how a more specific consent interface would increase auditability, which would help with getting recourse. P11 mentioned how knowing the extent of reversibility would make workers feel more in control of their data, and P5 wanted the employer to perform risk assessments for worst-case scenarios and to clearly communicate risks.

\para{Enact regulations to protect workers' rights \& privacy.} In addition to internal accountability mechanisms within the organization, participants called for stronger regulations to protect workers’ rights \& privacy, given their inability to rely on private entities to protect their rights. P12 felt ``regulation is the right approach because I do know that employers don't like getting sued.'' However, others critiqued the effectiveness of regulation, with P14 noting, after reviewing the regulation intervention sketch, that ``[a]ll of these are fine, except do we truly trust these?'' P13 was doubtful the company would comply, as ``companies lie a lot, all the time.'' Participants also desired stronger stipulations that ban selling employee health data, and wanted employee rights to be expanded in accordance with the ``reversible'' dimension of FRIES. Another example was guaranteeing employees the ``right to be forgotten'' after their employment ended, where they could request all of their data to be erased.
While a common suggestion is to aggregate data and make raw data unavailable, P12 argued that anonymizing and aggregating data is not sufficient to protect people's privacy. Further, P8 suggested that guaranteeing workers with healthcare independent of the employer can support more meaningful consent.

\para{Inform people of their basic workplace rights.}
Participants emphasized the need to raise awareness about workplace rights. Some wanted employers to hold
``a workshop on that, or have an educational stream to promote or have a conversation like starting a dialogue with the employees''~[P5]. Others like P12 were however skeptical about how trustworthy employers are and if they would be candid with the workers. This led them further emphasize the need for an external organization to help educate the workers, with P13 adding that placing the burden on other employees to educate their co-workers was unfair, exhausting, and impractical.

\section{Discussion and Conclusions}\label{section:discussion}
Our work cuts across scholarships on quantified workplace~\cite{mashhadi2016let,mirjafari2019differentiating,DasSwain2019FitRoutine}, privacy surrounding pervasive systems~\cite{barocas2014big,luger2014framework,jo2020lessons}, and feminist approaches to consent~\cite{parenthood2020sexual,strengers2021embodied,friedman2019yes}. We systematically identified both challenges to supporting meaningful consent for workplace wellbeing technologies (\autoref{sec:challenges}) and possible sociotechnical interventions (\autoref{sec:interventions}) to address them (\autoref{tab:mapping} provides a mapping).
In doing so, we reframe problems typically framed as individual and technical (like consent) into structural problems requiring sociotechnical solutions. We hope our work is a step toward a more consentful culture. 

\aptLtoX[graphic=no,type=env]{\begin{table}
\caption{Mapping of challenges \& interventions to support meaningful consent and FRIES dimensions. Interventions are color-coded: \hlPur{\textbf{Tech's Aims and Affordances}} and \hlPol{\textbf{Policies and Practices}} around the tech.}
\label{tab:mapping}
\begin{imageonly}
\begin{tabular}{@{}ll@{}c@{}}
\textbf{Challenge} & \textbf{Intervention} & \textbf{FRIES}\\ 
\hline
\multicolumn{3}{l}{\cellcolor{tablerowcolor2}\textbf{Facets of inherent power differential in the workplace}}\\
\multirow{3}{*}{\parbox{0.27\columnwidth}{Little (data) privacy guarantee at work}} & \cellPol{Ensure oversight from a public interest organization.} & F, R\\
\cdashlinelr{2-3}
& \cellPol{Enact regulations to protect workers’ rights and privacy.} & F, R\\
\hdashline
\multirow{5}{*}{{The workplace bottom line takes precedence over workers’ interests.}} & \cellPur{The technology should have a clear and sensible goal} & F, I\\
\cdashlinelr{2-3}
& \cellPur{The technology’s benefits should be commensurate with the cost of data sharing} & F\\
\cdashlinelr{2-3}
& \cellPur{Prioritize the interests of workers above business.} & F\\
\cdashlinelr{2-3}
& \cellPol{Create space for negotiations centering workers’ interests} & F, I\\
\cdashlinelr{2-3}
& \cellPol{A commitment from the company to do the right thing} & F, I\\
\hdashline
\multirow{4}{*}{{Employers have a high level of control over workers’ needs.}} & \cellPol{Show a strong record of trustworthy behavior.} & F\\
\cdashlinelr{2-3}
& \cellPol{Support recourse \& workers interrogating data practices} & R, S\\
\cdashlinelr{2-3}
& \cellPol{A commitment from the company to do the right thing} & F, I\\
\cdashlinelr{2-3}
& \cellPol{Enact regulations to protect workers’ rights and privacy} & F, R\\
\rowcollight \multicolumn{3}{l}{\textbf{Consequences of the power differential}}\\
\multirow{2}{*}{{Lack of accountability}} & \cellPol{Ensure oversight from a public interest organization.} & F, R\\
\cdashlinelr{2-3}
& \cellPol{Support recourse \& workers interrogating data practices.} & R, S\\
\hdashline
\multirow{2}{*}{{Information asymmetry about workplace rights and data use.}} & \cellPol{Inform people of their basic workplace rights.} & I\\
& \cellPol{} & \\
\cdashlinelr{2-3}
& \cellPol{Be transparent about who has access to workers’ data.} & I, S\\
& \cellPol{} & \\
\hdashline
\multirow{4}{*}{{Workers may not be able to negotiate.}} & \cellPol{Support recourse \& workers interrogating data practices.} & R, S\\
\cdashlinelr{2-3}
& \cellPol{Create space for negotiations centering workers’ interests.} & F, I\\
\cdashlinelr{2-3}
& \cellPur{The technology should facilitate an active reflection and adjustment of consent preferences.} & I, R\\
\cdashlinelr{2-3}
& \cellPur{Build interfaces that enable workers to better voice nuanced choices.} & R, S\\
\hdashline
\multirow{4}{*}{{Workers may feel coerced into consenting.}} & \cellPol{Support recourse \& workers interrogating data practices.} & R, S\\
\cdashlinelr{2-3}
& \cellPol{Do not assume consent in the workplace.} & F\\
\cdashlinelr{2-3}
& \cellPol{Encourage deliberation in the consent process to preserve workers’ autonomy.} & F, I\\
\cdashlinelr{2-3}
& \cellPur{Build interfaces that make saying no easy (and valid)} & R\\
\rowcollight \multicolumn{3}{l}{\textbf{Inherent risks posed by the technology}}\\
\multirow{2}{*}{{Sensing tech. often collect sensitive data.}} & \cellPol{Provide clarity about data governance practices.} & I, S\\
\cdashlinelr{2-3}
&\cellPol{Regulation to ensure strong data security practices.} & F, I\\
\hdashline
\multirow{2}{*}{{The technology may not be trustworthy.}} & \cellPur{Provide evidence and information about the technology’s efficacy, limits, and risks.} & I\\
\cdashlinelr{2-3}
&\cellPol{Show a strong record of trustworthy behavior.} & F\\
\rowcollight \multicolumn{3}{l}{\textbf{Technological barriers to meaningful consent}}\\
\multirow{3}{*}{{Technology may not support ongoing consent.}} & \cellPol{Encourage deliberation in the consent process to preserve workers’ autonomy.} & F, I\\
\cdashlinelr{2-3}
& \cellPur{Build interfaces that enable workers to better voice nuanced choices.} & R, S\\
\cdashlinelr{2-3}
& \cellPur{The technology should facilitate an active reflection and adjustment of consent preferences.} & I, R\\
\hdashline
\multirow{2}{*}{{Information asymmetry about the technology.}} & \cellPur{Provide evidence and information about the technology’s efficacy, limits, and risks.} & I\\
\cdashlinelr{2-3}
& \cellPur{The technology should facilitate an active reflection and adjustment of consent preferences.} & I, R\\
\hdashline
\multicolumn{1}{l}{Cannot say no without losing access to the technology.} & \cellPur{The technology should facilitate functionality-based consent.} & S\\
\hline
\end{tabular}
\end{imageonly}
\end{table}}{\begin{table}[t]
\sffamily
\footnotesize
\centering
\caption{Mapping of challenges \& interventions to support meaningful consent and addressed FRIES dimension. Interventions are color-coded: Tech's \hlPur{\textbf{Aims and Affordances}} and \hlPol{\textbf{Policies and Practices}} around the tech.}
\vspace{-6pt}
\label{tab:mapping}
\renewcommand{\arraystretch}{0.65}
\setlength{\tabcolsep}{1pt}
\resizebox{\columnwidth}{!}{\begin{tabular}{p{0.27\columnwidth}p{0.67\columnwidth}@{}c@{}}
\textbf{Challenge} & \textbf{Intervention} & \textbf{FRIES}\\ 
\hline
\multicolumn{3}{l}{\cellcolor{tablerowcolor2}\textbf{Facets of inherent power differential in the workplace}}\\
\multirow{3}{*}{\parbox{0.27\columnwidth}{Little (data) privacy guarantee at work}} & \cellPol{Ensure oversight from a public interest organization.} & F, R\\
\cdashlinelr{2-3}
& \cellPol{Enact regulations to protect workers’ rights and privacy.} & F, R\\
\hdashline
\multirow{5}{*}{\parbox{0.27\columnwidth}{The workplace bottom line takes precedence over workers’ interests.}} & \cellPur{The technology should have a clear and sensible goal} & F, I\\
\cdashlinelr{2-3}
& \cellPur{The technology’s benefits should be commensurate with the cost of data sharing} & F\\
\cdashlinelr{2-3}
& \cellPur{Prioritize the interests of workers above business.} & F\\
\cdashlinelr{2-3}
& \cellPol{Create space for negotiations centering workers’ interests} & F, I\\
\cdashlinelr{2-3}
& \cellPol{A commitment from the company to do the right thing} & F, I\\
\hdashline
\multirow{4}{*}{\parbox{0.27\columnwidth}{Employers have a high level of control over workers’ needs.}} & \cellPol{Show a strong record of trustworthy behavior.} & F\\
\cdashlinelr{2-3}
& \cellPol{Support recourse \& workers interrogating data practices} & R, S\\
\cdashlinelr{2-3}
& \cellPol{A commitment from the company to do the right thing} & F, I\\
\cdashlinelr{2-3}
& \cellPol{Enact regulations to protect workers’ rights and privacy} & F, R\\
\rowcollight \multicolumn{3}{l}{\textbf{Consequences of the power differential}}\\
\multirow{2}{*}{\parbox{0.27\columnwidth}{Lack of accountability}} & \cellPol{Ensure oversight from a public interest organization.} & F, R\\
\cdashlinelr{2-3}
& \cellPol{Support recourse \& workers interrogating data practices.} & R, S\\
\hdashline
\multirow{2}{*}{\parbox{0.25\columnwidth}{Information asymmetry about workplace rights and data use.}} & \cellPol{Inform people of their basic workplace rights.} & I\\
& \cellPol{} & \\
\cdashlinelr{2-3}
& \cellPol{Be transparent about who has access to workers’ data.} & I, S\\
& \cellPol{} & \\
\hdashline
\multirow{4}{*}{\parbox{0.25\columnwidth}{Workers may not be able to negotiate.}} & \cellPol{Support recourse \& workers interrogating data practices.} & R, S\\
\cdashlinelr{2-3}
& \cellPol{Create space for negotiations centering workers’ interests.} & F, I\\
\cdashlinelr{2-3}
& \cellPur{The technology should facilitate an active reflection and adjustment of consent preferences.} & I, R\\
\cdashlinelr{2-3}
& \cellPur{Build interfaces that enable workers to better voice nuanced choices.} & R, S\\
\hdashline
\multirow{4}{*}{\parbox{0.25\columnwidth}{Workers may feel coerced into consenting.}} & \cellPol{Support recourse \& workers interrogating data practices.} & R, S\\
\cdashlinelr{2-3}
& \cellPol{Do not assume consent in the workplace.} & F\\
\cdashlinelr{2-3}
& \cellPol{Encourage deliberation in the consent process to preserve workers’ autonomy.} & F, I\\
\cdashlinelr{2-3}
& \cellPur{Build interfaces that make saying no easy (and valid)} & R\\
\rowcollight \multicolumn{3}{l}{\textbf{Inherent risks posed by the technology}}\\
\multirow{2}{*}{\parbox{0.25\columnwidth}{Sensing tech. often collect sensitive data.}} & \cellPol{Provide clarity about data governance practices.} & I, S\\
\cdashlinelr{2-3}
&\cellPol{Regulation to ensure strong data security practices.} & F, I\\
\hdashline
\multirow{2}{*}{\parbox{0.25\columnwidth}{The technology may not be trustworthy.}} & \cellPur{Provide evidence and information about the technology’s efficacy, limits, and risks.} & I\\
\cdashlinelr{2-3}
&\cellPol{Show a strong record of trustworthy behavior.} & F\\
\rowcollight \multicolumn{3}{l}{\textbf{Technological barriers to meaningful consent}}\\
\multirow{3}{*}{\parbox{0.25\columnwidth}{Technology may not support ongoing consent.}} & \cellPol{Encourage deliberation in the consent process to preserve workers’ autonomy.} & F, I\\
\cdashlinelr{2-3}
& \cellPur{Build interfaces that enable workers to better voice nuanced choices.} & R, S\\
\cdashlinelr{2-3}
& \cellPur{The technology should facilitate an active reflection and adjustment of consent preferences.} & I, R\\
\hdashline
\multirow{2}{*}{\parbox{0.25\columnwidth}{Information asymmetry about the technology.}} & \cellPur{Provide evidence and information about the technology’s efficacy, limits, and risks.} & I\\
\cdashlinelr{2-3}
& \cellPur{The technology should facilitate an active reflection and adjustment of consent preferences.} & I, R\\
\hdashline
\multicolumn{1}{p{0.25\columnwidth}}{Cannot say no without losing access to the technology.} & \cellPur{The technology should facilitate functionality-based consent.} & S\\
\hline
\end{tabular}}
\vspace{-10pt}
\end{table}}

\para{Adapting FRIES for workplace wellbeing tech.}
Our work draws on FRIES to guide discussions on how to support meaningful consent to passive sensing in the workplace. Our participants envision varying ways to support different dimensions of FRIES. For instance, several suggested interventions---such as ensuring that consent is not assumed and that there are 
no penalties for saying no---support the ``freely-given'' dimension. However, ``enthusiasm'' was deemed more complex, as some critiqued it as both sometimes unnecessary (if the technology was valuable enough), and hard to evaluate in the workplace (where a manager's encouragement could lead to feeling coerced and compromise genuine enthusiasm).

We also found that the workplace context amplifies two implicit aspects in FRIES: \textit{accountability} and \textit{reciprocity}, which participants considered essential for meaningful consent. 
Participants described how getting commensurate benefits to the cost of sharing data and accountability from the employer would increase enthusiasm. These dimensions fit well into FRIES by centering relationality on how one can expect to be treated by the other party. \citeauthor{im2021yes} applied FRIES to user-to-peer interactions on social platforms and user-to-system interactions. Likewise, for workplace technologies, FRIES can be expanded to consider employee-to-employer interactions, which introduce the greater need for accountability and reciprocity. Our results emphasize the importance of regulations to ensure greater transparency in informing workers about their rights, data governance, and the technology's efficacy---supporting multiple dimensions of FRIES. Relatedly, while FRIES primarily supports ``affirmative consent'' and guides design choices to increase user autonomy to say ``yes,'' multiple participants advocated for protecting the right to say ``no,'' given the coercive forces in the workplace that place workers under pressure to say ``yes;'' with some participants thus also expressing skepticism about workers being able to meaningfully consent to workplace sensing tech.

\para{Cultivating a trust-based relationship with workers.}
Prior work argued that consent in the workplace should be framed as an organizational behavior issue, not just a legal one~\cite{bohns2020consent}. Relatedly, our participants also underscored that establishing trust between employers and employees is critical for consent. But how exactly do we establish this trust? Our focus groups surfaced interventions that could help build trust, including a clear and sensible purpose for the technology, increased transparency in communication, prioritizing the interests of workers, and creating space for interrogation and negotiation, aligning with recommendations from prior work~\cite{bohns2020consent,maltseva2020wearables,andrus2021we,norval2022disclosure}. Participants thought fostering a trust-based relationship and centering worker interests are fundamental to being a ``benevolent employer.'' Prior work also proposed making employers more beholden to worker interests, such as designating them as employee information fiduciaries~\cite{bodie2022employers} or instituting an internal privacy due diligence process~\cite{ebert2021big}. However, other participants critiqued the idea of the ``benevolent employer'' because the inherent power asymmetry leads workers to naturally distrust their employer~\cite{introna2000workplace}. Our participants suggested stronger regulation to protect workers' rights as an alternative way to support consent---also corroborated by worker data protection literature~\cite{kovacs2020informed, consumerreports2022}.

\para{Is meaningful consent to workplace wellbeing sensing technologies possible?}
Participants questioned the very use of these technologies within the workplace. P12 rejected ``the idea of having my workplace know anything about my health [..] even if I consent, I reject the premise of this system.'' Prior work also noted privacy harms from the use of passive sensing technologies in the workplace~\cite{corvite2022data,crawford2021time,heaven2020faces,kappas2010smile}. We also situate these findings with the design refusal literature~\cite{grae2020responsibility}, or rejecting the premise of these systems entirely, including recent calls to ban the use of these technologies~\cite{consumerreports2022, kovacs2020informed, crawford2021time}. Our participants' concerns about the validity of these technologies also corroborate prior work~\cite{docherty2022re, roemmich2021data, barrett2019emotional, adler2022burnout, stark2021ethics} and raise questions about whether these technologies, which aim to know an individual better than they may know themselves~\cite{kaur2022didn}, can satisfy fundamental privacy conditions, such as ``semantic discontinuity''~\cite{cohen2019turning}---the right to be somewhat unknown and undefined.

Our work finds several barriers that must be overcome before workers can meaningfully consent to workplace wellbeing technologies. While this paper is not the solution to these barriers, it highlights several considerations that would need to be incorporated into the technology and the sociotechnical ecosystem surrounding it. The feasibility and efficacy of these potential interventions are, however, still unclear and hard to foresee. For instance, participants described how technological interventions that provide more flexibility and control---by changing how and how often workers provide consent---can have their own unintended consequences. P12 noted how requiring more interactions to provide consent could lead to fatigue and eventually meaningless consent:
``Everyone thought that the cookie consent stuff in GDPR would be great, but the more times you ask people things, the more they get fatigued, and they just end up either disagreeing or agreeing to everything.'' P15 similarly noted that users could become overwhelmed by having to proactively go back and review all data agreements. Similar issues with ``consent fatigue'' have also been noted in prior work~\cite{schermer2014crisis,teare2021reflections}. Our focus groups also surfaced that there may be more suitable ways of supporting workplace wellbeing than implementing and relying on wellbeing sensing tech. P12 suggested they would prefer if they were instead given a health stipend to use as they wished. Overall, companies should prioritize workers' interests, autonomy, and privacy in deploying workplace wellbeing technologies. This is consistent with literature finding that preserving workers' privacy improves their wellbeing and organizational outcomes~\cite{maltseva2020wearables, bohns2020consent}. 

\para{Limitations \& Future Directions.} We explored consent for a hypothetical (but theoretically feasible) technology, primarily to surface a broader set of challenges without being constrained by the nitty-gritty of a particular technology. 
While this enables us to map a wider range of issues, it may also make it harder to surface issues arising in specific deployment settings. Future work should examine specific technologies to gather complementary insights. Adopting a studying-up approach~\cite{barabas2020studying} to examine the perspectives of relevant decision-makers would also complement our findings. 
While we recruited participants to cover a diversity of roles and perspectives with respect to consenting to workplace technologies---most of our participants are from so-called `WEIRD' societies (western, educated, industrialized, rich, and democratic)~\cite{henrich2010most}. Participants were all white or Asian; highly educated; with expertise in software, privacy, or AI; and a majority also work in tech. Their views will naturally reflect their expertise and experiences and might thus not be representative of the views of more typical workers who might be much less knowledgeable about data systems or AI. Future work focusing on non-tech workers would complement the set of challenges and possible interventions we identified.       

This study also focused on the interface/interactions when consent decisions are made as this is the entry point and is most easily represented. However, our findings suggest that the technology and consent practices need to be holistically changed, and 
future work can consider a more holistic lens like the privacy-centeredness~\cite{giannopoulou2020algorithmic} and seamfulness~\cite{vertesi2014seamful} of technologies.




\begin{acks}
This research was conducted at Microsoft Research. We thank David Widder and other participants for making time for the study. We thank Q. Vera Liao, Ziang Xiao, Samir Passi, Sachin Pendse, Shamsi Iqbal, Javier Hernandez, Mary Czerwinski, Kate Crawford, and Solon Barocas for providing feedback and resources for this work. 
\end{acks}

\bibliographystyle{ACM-Reference-Format}
\bibliography{source/0paper}
\balance{}

\appendix
\clearpage
\section{Appendix}

\setcounter{table}{0}
\setcounter{figure}{0}
\renewcommand{\thetable}{A\arabic{table}}
\renewcommand{\thefigure}{A\arabic{figure}}

\begin{figure}[h!]
\centering
    \begin{subfigure}[b]{0.95\columnwidth}
    \centering
    \includegraphics[width=\columnwidth]{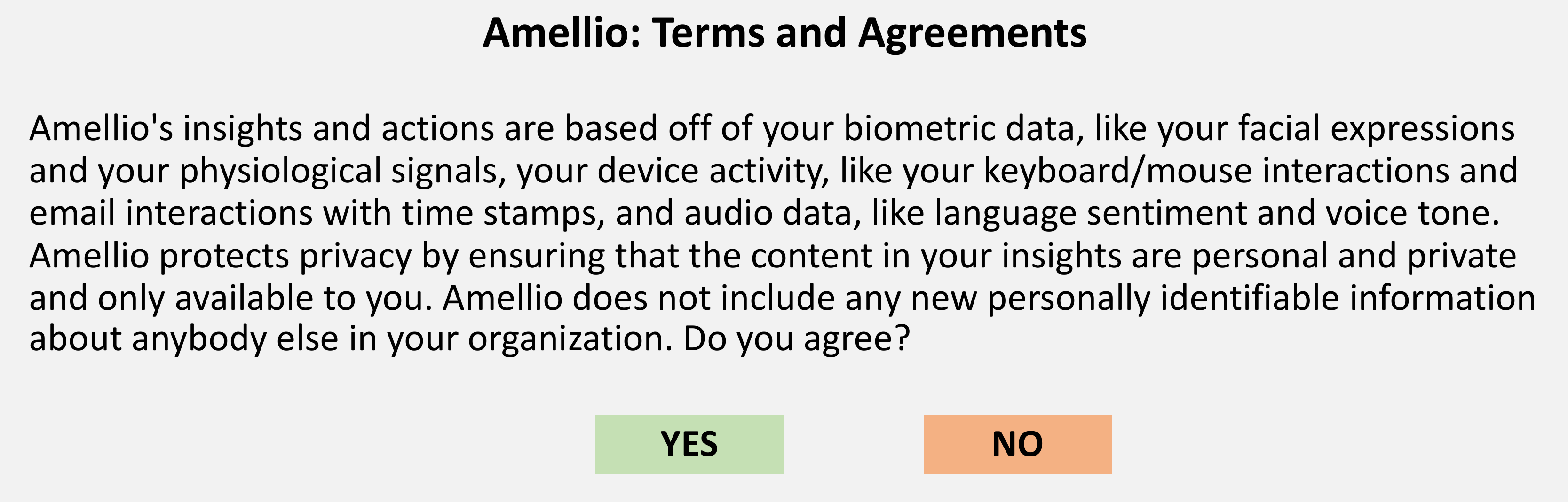}
    \caption{Terms and Conditions}
    \label{fig:amellio_terms}
    \end{subfigure}\hfill
    \begin{subfigure}[b]{0.95\columnwidth}
    \centering
    \includegraphics[width=\columnwidth]{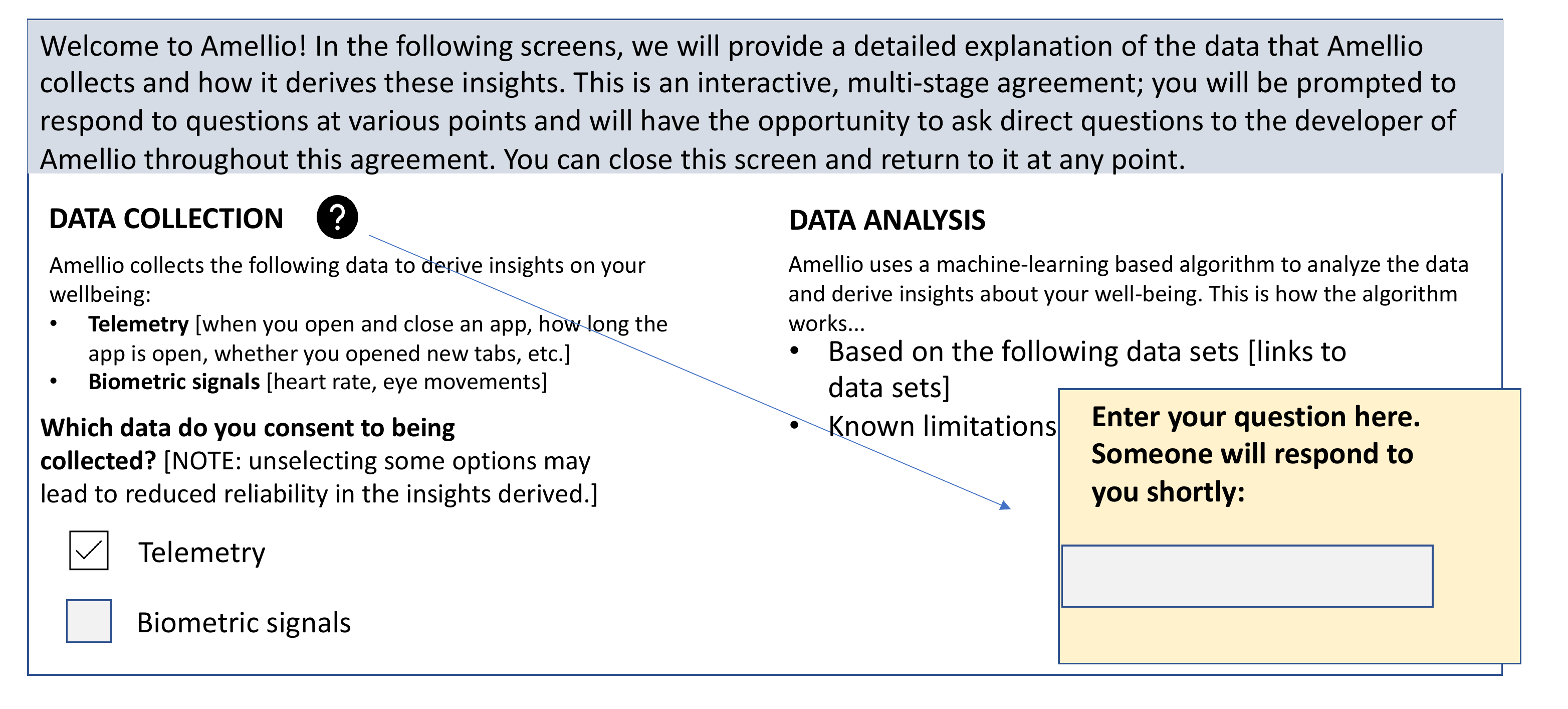}
    \caption{Interface re-design}
    \label{fig:intervention1}
    \end{subfigure}\hfill
    \vspace{6pt}
    \begin{subfigure}[b]{0.95\columnwidth}
    \centering
    \includegraphics[width=\columnwidth]{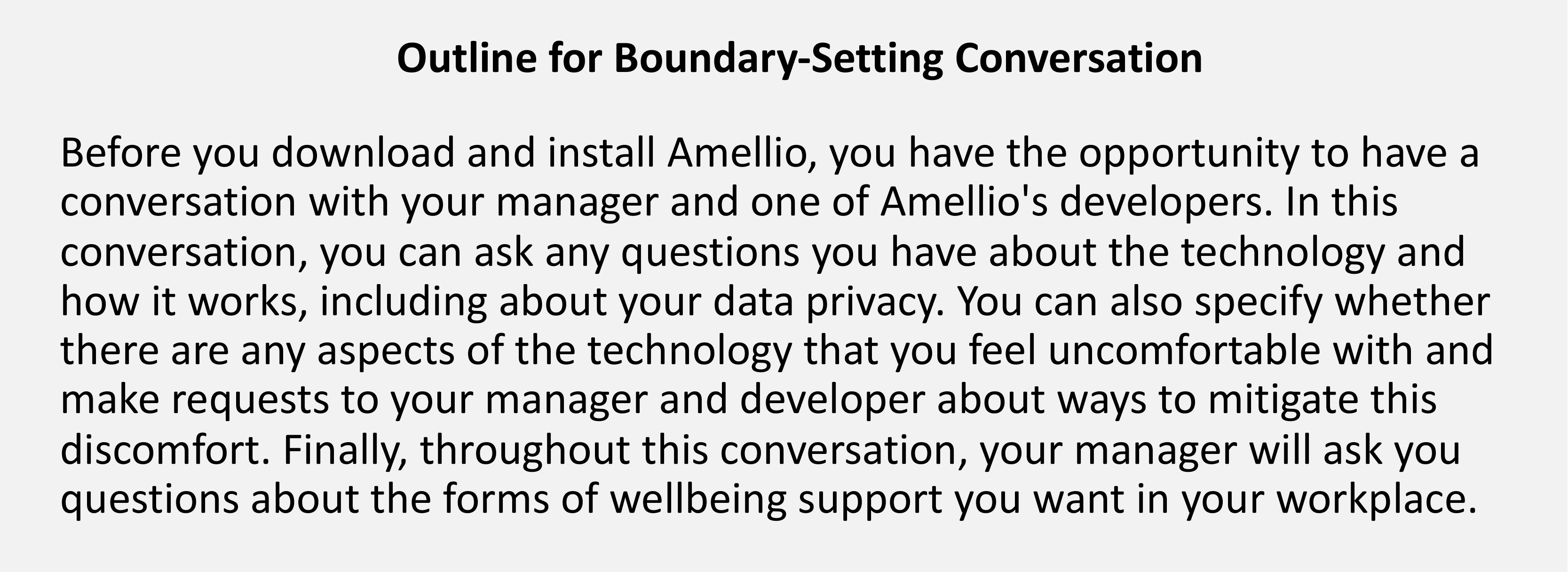}
    \caption{Organizational practice}
    \label{fig:intervention2}
    \end{subfigure}\hfill
        \begin{subfigure}[b]{0.95\columnwidth}
    \centering
    \includegraphics[width=\columnwidth]{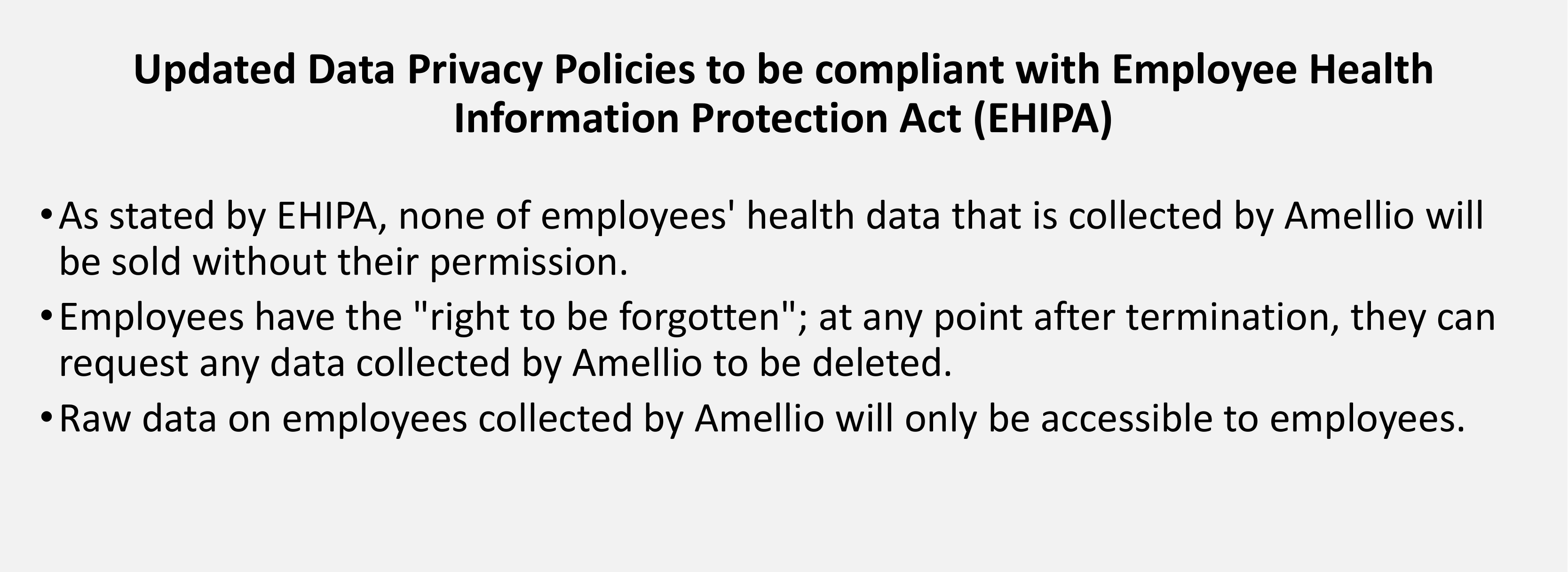}
    \caption{Data protection regulation}
    \label{fig:intervention3}
    \end{subfigure}\hfill
    \caption{Sketches shown to participants on the hypothetical workplace wellbeing sensing technology, Amellio, (a) Amellio's initial terms and agreement; (b), (c), (d) interventions to help participants think about what could support meaningful consent to Amellio.}
    \Description[figure]{Four figures describing the prompts shown to the participants about the hypothetical workplace wellbeing sensing technology (Amellio), and the interventions to help support meaningful consent. Figure (a) has the initial terms and conditions of Amellio; Figure (b) includes a re-designed consent interface where someone can select specific options and also seek more details about the data being collected; Figure (c) includes an outline for boundary-setting conversations as an intervention; Figure (d) includes an intervention through an updated data protection regulation.}
\end{figure}




\end{document}
\endinput